# Effect of TiO$_2$ nanoparticles on the thermal stability of native DNA under UV irradiation

Evgeniya Usenko[a], Alexander Glamazda[b], Vladimir Valeev , and Victor Karachevtsev

B.Verkin Institute for Low Temperature Physics and Engineering of the National Academy of Sciences of Ukraine, 47 Nauky Ave., Kharkiv 61103, Ukraine

[a] usenko@ilt.kharkov.ua, [b] glamazda@ilt.kharkov.ua

**Abstract**

TiO$_2$ nanoparticles (NPs) are widely used in the environmental engineering, medicine, chemical and food industries due to their unique photocatalytic and biocidal properties. NPs may generate reactive oxygen species and, hence, have the toxic effect on the living cells via oxidative stress. An external UV irradiation may magnify the photocatalytic properties of TiO$_2$ NPs. In this regard, we have analyzed the influence of TiO$_2$ NPs on the conformation and thermal stability of native DNA in a buffer solution without and under UV irradiation exploiting absorption spectroscopy with DNA thermal denaturation in the range of 20-94$^0$C. Upon DNA heating from about 25 to 44$^0$C, we have observed the stabilization of DNA duplex in the presence of TiO$_2$ NPs. This additional biopolymer stabilization indicates that partial DNA unwinding appears as a result of the direct binding of the biopolymer to NPs. We showed that the performed UV treatment of DNA during 3 hours leads to partial unwinding of the biopolymer structure. The NPs injection to the biopolymer solution induced the additional effect on the DNA thermal stability under UV irradiation. The performed analysis of the experimental data suggests that the nature of the impact of NPs on the biopolymer is complex.

## 1 Introduction

Recently, in nanoscience, considerable interest has been paid to the study of the problems of the effect of nanomaterials on biological objects. First of all, this is caused by a breakthrough in the synthesis and the development of the research methods for the nanoscale materials. In the nanoobjects, due to the confine geometry of the propagation of quantum excitations, new physical properties are generated which are not characteristic for the bulk materials. Among such materials, TiO$_2$ nanoparticles (NPs) occupy an important place. They are chemically inert, inexpensive, besides they have high thermal stability and high photocatalytic activity that can be revealed even under sunlight. The latter factor can be enhanced by absorbing the focused UV irradiation. Under ultraviolet exposure, electrons in the valence band of TiO$_2$ NPs absorb the light with energy equal or more than the energy gap between the valence and conduction bands.

It induces generating the electron-hole pairs: the excited electron from the valence band jumps to the conduction band, leaving the hole in the valence band. Both photoexcited electrons can participate in the reduction of oxygen and the holes can oxidize the water molecules adsorbed on the photocatalytic surface of NPs. The photoexcited electron can react with oxygen generating the reactive oxygen species (ROS), such as hydroxyl (OH) and peroxy ($HO_2$) radicals, superoxide anions ($O_2^-$), and hydrogen peroxide ($H_2O_2$) which take part, for example, in the degradation of the organic pollutants. It makes $TiO_2$ nanoparticles attractive for antibacterial and self-cleaning surface coatings [1]. In addition, $TiO_2$ NPs might be used as the promising carriers for drug-delivery application and the sensitive nanosensor elements for $O_2$, $H_2O_2$, humidity, etc. [2-4]. Moreover, by the tuning of the size and shape of $TiO_2$ NPs, it is possible to increase their biological activity and high penetration ability into the body. Penetrating into the cell, nanoparticles are able to interact with the structural components of the cell, causing their damage [5]. Despite the large scale of production and extensive use of $TiO_2$ NPs in various areas of human activity, their toxic properties have not been sufficiently studied, although at the moment extensive studies are being conducted to evaluate their safety and potential risks [2, 6, 7-10]. In addition to $TiO_2$ NPs, DNA is also affected by other harmful factors, such as UV radiation. It is one of the most effective and carcinogenic exogenous factors that is able to affect DNA and alter the genome integrity. All of that, ultimately, can affect the normal life processes of all organisms ranging from prokaryotes to mammals [11]. At the moment, there are already a number of works devoted to the study of the physicochemical properties of NPs (see, e.g. [12, 13]), as well as the interaction of $TiO_2$ NPs with DNA (see, e.g. [14, 15]). In particular, in [16], the effect of $TiO_2$ NPs on DNA was studied using spectroscopic methods. The obtained data indicate the presence of electrostatic interaction along with the formation of the chemical bonds between $TiO_2$ NPs and DNA [16]. This binding induces the structural changes in the biopolymer itself. The molecular mechanisms of DNA damage into HepG2 cells induced by $TiO_2$ NPs were recently studied in Ref. [17]. There was shown that $TiO_2$ NPs are able to effect on the gene expression changes in DNA. Also, authors confirmed an activation of the elevated oxidative stress, including the generation of ROS, with increased hydrogen peroxide levels, decreased glutathione peroxidase, and reduced glutathione and activated caspase-3 levels in cells exposed to $TiO_2$ NPs. It was observed that $TiO_2$ NPs induce the expression of seventeen DNA damage marker genes [17]. Authors reported [5] the study including both chromosome aberration and comet assays at low dose exposure $TiO_2$ NPs. There was performed the study of the binding affinity with DNA using fluorescence titration method. This work also reported that NPs have a high affinity for DNA and can directly bind to one. In addition, it has been shown that NPs are able to strongly inhibit DNA replication and change the normal conformation of the polynucleotide that can lead to genotoxicity [5]. There was determinate the binding constant of $TiO_2$ NPs with DNA that is about ~ $4.2*10^6$ $M^{-1}$ [5]. Such a rather high value of the binding constant indicates a strong interaction between $TiO_2$ NPs and DNA. As mentioned above, other exogenous factors also have the destructive effect on DNA. One of them is UV radiation. It is known that UV radiation can lead to damage to DNA and important photosynthetic structures, a cellular damage or even cell death. And despite the fact that many living cells are able to repair simple damages. The enhanced exposure to ultraviolet radiation can cause irreversible damage. For example, in work [11], devoted to the study of the effect of UV radiation on DNA, it is reported that UV radiation is one of the powerful agents that can cause various mutagenic and cytotoxic DNA damage, its double-strand denaturation, disrupting the integrity of the genome.

At the moment, there is no data in the literature on the study of the temperature effect on the stability of DNA conformation in the presence of $TiO_2$ NPs and influence of the UV irradiation on the polymer stability in this surrounding. In this regard, the study of the $TiO_2$ NPs effect on natural DNA under UV irradiation is the topical issue of bioengineering. The aim of this study is obtaining the information on the $TiO_2$ NPs effect on the structural stability of DNA under an exposure to UV radiation. The interaction of $TiO_2$ NPs with DNA in a buffer solution in the presence and absence of UV irradiation was studied with UV spectroscopy and thermal denaturation in the range of 20-94$^0$C. It was revealed the addition of $TiO_2$ NPs to the buffer solution of DNA leads to the unfolding of its secondary structure. Heating of DNA:$TiO_2$ NPs assemblies in the temperature range of about 25-44$^0$C leads to the increase of a stability of the helical biopolymer conformation. We suppose that this effect may be due to the direct binding of DNA to $TiO_2$ NPs. It was also revealed that UV irradiation of DNA:$TiO_2$ NPs assemblies during 3 hours induces the partial denaturation of the biopolymer structure and shifts the temperature range of the increase of the stability of the helical conformation to higher temperatures (to about 38-66$^0$C). The analysis of the obtained experimental data suggests that the interaction of $TiO_2$ NPs with a biopolymer has complex nature and could be explained by both the direct and indirect effects of $TiO_2$ NPs on DNA. As it is known, the direct binding of $TiO_2$ NPs to DNA can occur in the following scenarios proposed in [16]: 1) the electrostatic interaction occurring between positively charged $TiO_2$ NPs and the negatively charged phosphate group of DNA; 2) the interaction of $TiO_2$ NPs with nitrogenous bases in the DNA groove; 3) the formation of the P-O-Ti covalent bonds between the phosphate backbone and the NPs surface. We suppose $TiO_2$ NPs can directly bind to guanine N7 atoms located in the DNA major groove by the coordination bond. The possibility of such interaction was shown for the example of DNA binding to AgNPs and AuNPs [18]. The indirect effect is related to the NPs generation of ROS inducing the damage of the DNA structure as a result of the UV exposure.

## 2 Materials

The powder of $TiO_2$ NPs (d<100 nm, $M_w$=79.87 g/mol) manufactured by Sigma (USA) was used in the present work. The $TiO_2$ NPs material was dispersed in distilled water and ultrasonicated with ν=22 kHz for 40 min at room temperature. The salmon sperm DNA ($M_w$=4*10$^6$-6*10$^6$ Da) purchased from Serva (Germany) was added to a buffer solution with 10$^{-3}$ M sodium cacodylate $(CH_3)_2AsO_2Na \cdot 3H_2O$ from Serva (Germany), 0.099 M NaCl at pH 5. The effect of $TiO_2$ NPs on the biopolymer structural conformation has been studied by adding the required amount of nanoparticles to the DNA buffer solution. To study the effect of UV radiation on the structural features of the biopolymer, the assemblies of $TiO_2$ NPs with DNA were irradiated with an ultraviolet light-emitting diode (λ=360-365 nm) for 3 hours (P=15 mW). The concentration of polynucleotide phosphates [P] equaled $(7.14\pm1)*10^{-5}$ M was determined by the molar extinction coefficient in the UV absorption maxima at $ν_m$=38500 см$^{-1}$ [19].

## 3 Methods

### 3.1 UV-spectroscopy

The absorption spectra in the visible and ultraviolet regions are caused by the transitions of the molecule from the ground to excited electronic states, accompanied by the absorption of the light with the frequency ν [20]. The light absorption within the ultraviolet range of 30000-50000 cm$^{-1}$ caused by π→π* and n→π* electronic transitions in the nucleobases of the biopolymer. In the DNA double helix, the nitrogenous bases stack one above the other. The change in the conformation and structural stability of nucleic acids occurs under the influence of various factors (for example, temperature, ionic conditions, UV exposure, etc.) and causes significant changes in the UV absorption spectrum [21]. The extinction coefficient of a double helix depends on mutual orientation of the intrinsic dipole moments of stacking nitrogenous bases. If the moments are collinear, the light absorption will decrease (hypochromism). If the moments are canted, the light absorption will increase (hyperchromism). In this work, the dependence of the optical density of solutions (A) of DNA with $TiO_2$ NPs on the wavenumber at room temperature (T=$T_0$=25°C) was recorded using UV-visible spectrophotometer (Specord M40, Carl Zeiss Jena, Germany).

### 3.2 Thermal denaturation

The performed thermal measurements on DNA are important for study of the possible structural biopolymer conformations and the development of the thermodynamic models. In the present work, thermal denaturation was used to study the structural stability of DNA in the presence of $TiO_2$ NPs. Upon increasing the temperature, the ordered structure of the polynucleotide undergoes the denaturation transition. As a result, the biopolymer forms loops which appear because of breaking some H-bonds between the nitrogen paired bases. When the temperature reaches a value close to the melting temperature ($T_m$), the number of the helical regions becomes equal to the number of the untwisted regions. In other words, $T_m$ is the temperature at which 50% of a DNA sequence is in the helix conformation, and the other 50% is present as single strands. A further increase in the temperature leads to a strong shift of equilibrium towards the increase of the fraction of the single strands. This process is accompanied by an increase of the intensity of UV absorption. The dynamics of this increase stops when the double helix of the biopolymer is completely untwisted. The melting curve describes the dependence of UV absorption intensity on temperature. The melting curves of the polynucleotides with different $TiO_2$ concentrations were recorded using UV-spectrophotometer at fixed wavenumber of $\nu_m=38500$ см$^{-1}$ that corresponds to the maximum absorption of DNA. We have used the laboratory software allowing to perform the registration of the melting curves as the temperature dependence of the hyperchromicity coefficient: $h(T)=[\Delta A(T)/A_{To}]_{\nu m}$, where $\Delta A(T)$ is a change in the optical density of DNA solution upon heating and $A_{To}$ is the optical density at $T=T_o$. Thus, $h(T)$ is the quantitative characteristic of hyperchromism. The registration of the absorption intensity was carried out using the double-cuvette scheme: identical solutions of polynucleotides or their complexes with $TiO_2$ were placed in both channels of the spectrophotometer. The reference cuvette was thermostated at $T=T_0\pm0.5°C$, while the sample cuvette was slowly heated at a rate of 0.25°C/min from 20 to 94°C.

## 4 Results and Discussion

### 4.1 Absorption spectroscopy of $TiO_2$ NPs

Recently [22,23], it was shown that one of the main factors determining the biological (and toxic) properties of NPs is their charge. It is known [24], that $TiO_2$ NPs are positively charged at low pH (pH<6.5) and become negative at high pH (pH>6.5). NPs are electrically neutral at pH~6.5. Moreover, in the pH range of about 2-5 and 9-12, the value of zeta-potential (ξ) displays a maximum in its absolute value. It means that the NPs suspension is most stable at these pH values [24]. The colloids with a higher value ξ are electrically more stable in comparison to colloids having a lower value of this parameter [24]. It was noted that the biological effect of positively charged nanoparticles has more toxicity than negatively charged ones [22]. It manifests itself in the destruction of the cell membranes. This effect is due to the fact that the positively charged nanoparticles penetrate the cells better than the negatively charged ones, both in the serum-free and serum-supplemented media [22]. In addition, the geometrical characteristics should also be taken into consideration in the estimation of the hazard of NPs

[23]. The smaller the size of NPs, the stronger the toxic effects were manifested. This is due to the fact that the decrease in the size of NPs leads to an increase in the area of the active surface. With a comparative estimation of the toxicity of nano- and micrometer sizes of the particles, it is known that the first, mainly, has more pronounced damaging effects. For example, such dependence has been revealed under comparative studies of TiO$_2$ NPs having a size of 20-30 nm in a comparison with particles of 200-250 nm [23]. The TiO$_2$ nanoparticles used within the present work previously have been identified with UV-Vis absorption spectroscopy. Fig.1 shows the absorption spectrum of TiO$_2$ NPs dissolved in aqueous solution. We have focused on the low-energy part of UV spectra of TiO$_2$ particles with maxima at about 340 nm that can be mainly attributed to the ligand-to-metal charge transfer $O^{2-} \rightarrow Ti^{4+}$ ($O_{2p} \rightarrow Ti_{3d}$). The band gap energy (eV) of our sample was calculated using the following equation: $E_g = hc/\lambda_g$, where $\lambda_g$ is the absorption onset wavelength (nm) of the exciting light, $c$ is velocity of light and $h$ is Planck's constant. The band gap energy value calculated using the presented equation was evaluated as 2.94 eV. The recorded absorption spectrum of TiO$_2$ NPs corresponds to the calculated data presented in [25] for nanoparticles with a primary d≤100 nm.

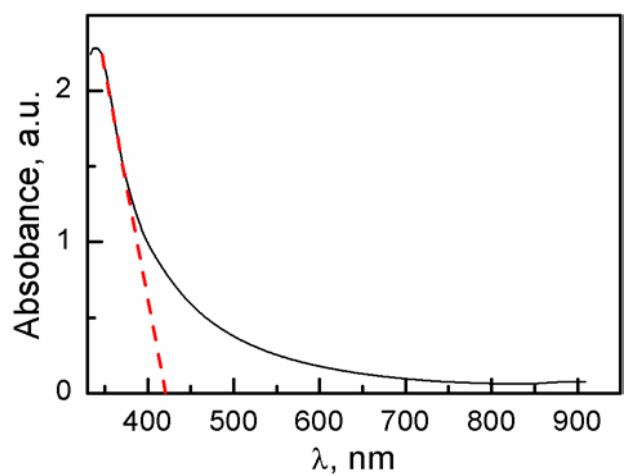

**Fig.1** UV-Visible absorption spectrum of TiO$_2$ NPs. The absorption onset wavelength ($\lambda_g$) is obtained through extrapolation (the red dash line) of the linear part of the spectrum.

### 4.2 Study of DNA conformation in the presence of TiO$_2$ NPs

The typical temperature dependence of hyperchromic coefficient (h) of DNA has the usual S-like shape: heating induces the helix-coil structural transition, as reflected in the appearance of the hyperchromism (h>0) (Fig. 2, curve 1). The analysis of the DNA melting curve without TiO$_2$ NPs allows us to determine the thermodynamic parameters of the biopolymer: $T_m = 78.5\,^0C$ and $h_{m0} = 0.41$ (Fig. 2) (where $h_{m0}$ - the value of the hyperchromic coefficient in the absence of TiO$_2$ NPs). However, the addition of TiO$_2$ NPs to the DNA buffer solution causes the shape of the biopolymer melting curve changes dramatically. Upon heating the trough appears in the temperature range of about 30-75 $^0$C. The depth of the trough increases with the TiO$_2$ concentration ([$c_{TiO2}$]). As [$c_{TiO2}$] increases, several segments can be clearly distinguished in the melting curve. Fig. 2 shows the example of the decomposition of the DNA melting curve in the presence of TiO$_2$ with [$c_{TiO2}$]=1.5*10$^{-4}$M. The characteristic points: $T_{s1}$ and $T_{s2}$ as well as $T_{f1}$ and $T_{f2}$ were calculated using an extrapolation of the linear segments of the melting curves. They are connected to the start ($T_{s1}$, $T_{s2}$) and finish ($T_{f1}$, $T_{f2}$) of the definite ongoing processes discussed below. In the segment of a-b, with an increase in temperature, absorption hypochromism is observed, the value of which increases with [$c_{TiO2}$]. We believe that this effect is caused by the manifestation of the additional stabilization of the biopolymer structure in the presence of TiO$_2$ NPs. As it was mentioned above, one of the mechanisms of TiO$_2$ binding to DNA is the electrostatic interaction of positively charged TiO$_2$ NPs with negatively charged phosphate groups of DNA [16]. It is known that heating leads to the formation of the unwound regions at the ends as well as in the helix segments located far from the biopolymer ends. The unwound regions can effectively interact with NPs resulting in the stabilization of the biopolymer from the melting. In the segment of b-c, the DNA ordered structure remains stable (the value of h does not change). The destruction of the stable biopolymer structure occurs in the segment of c-d and it ends with the plateau (Fig. 2).

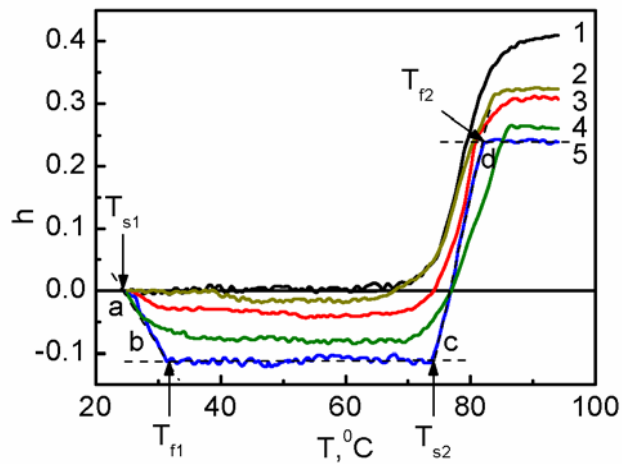

**Fig. 2** Temperature dependence of hyperchromic coefficient of DNA without (curve 1) and with (curves 2-5) TiO$_2$ NPs: «1» - [$c_{TiO2}$]=0; «2» - [$c_{TiO2}$]=2.5·10$^{-5}$ M; «3» - [$c_{TiO2}$]=5·10$^{-5}$ M; «4» - [$c_{TiO2}$]=10$^{-4}$ M; «5» - [$c_{TiO2}$]=1.5·10$^{-4}$ M. The extrapolation of the linear segments is shown in the black dashed lines.

### 4.3 UV-irradiation effect on DNA conformation with TiO$_2$ NPs

As mentioned above, one of the known exogenous factors having the destructive effect on DNA structure is UV radiation that is able to induce the appearance of as the lope as the single-stranded fragments in the biopolymer. Here we describe the spectroscopic studies of UV irradiated DNA:TiO$_2$ assemblies. Fig. 3 shows the temperature dependence of the hyperchromic coefficient of DNA obtained without and in the presence of TiO$_2$ NPs under UV irradiation that was carried out during 3 hours. The melting curves for non-irradiated and irradiated DNA solutions demonstrate the different dynamics. The melting temperature of DNA decreases to more than 6$^0$C and the h$_{m0}$ is equal to 0.35 under UV irradiation for three hours. This effect may be caused by the partial denaturation of DNA and the appearance of single-stranded fragments under the UV irradiation [11,26]. The formation of single-stranded fragments in the biopolymer in the complexes with TiO$_2$ NPs under the UV irradiation is confirmed by the presence of so-called non-cooperative "tails" in DNA melting curves (see, the a'-a segment in Fig. 3). The appearance of less cooperative melting segments on irradiated DNA:TiO$_2$ NPs melting curves (Fig. 3) (in comparison with the melting curves of the non-irradiated biopolymer assemblies presented in Fig. 2) is due to the fact that the melting of single-stranded DNA fragments formed as a result of UV irradiation of the native biopolymer contributes to the total dependence of the hyperchromic coefficient of double-stranded DNA. With increasing [c$_{TiO2}$], a trough appears in the melting curves in the temperature range of 60–75$^0$C (Fig. 4). It should be noted that the temperature of the onset (T$_{s1}$) and end (T$_{f1}$) of the manifestation of the structural stabilization of irradiated DNA shifts to the high-temperature region relative to the melting curve of non-irradiated DNA (Figs. 3, 4). So, for example, at [c$_{TiO2}$]=1.5*10$^{-4}$M, T$_{s1}$=25.3$^0$C and T$_{f1}$=31.4$^0$C in the absence of irradiation, and T$_{s1}$=41.6$^0$C and T$_{f1}$=64.8$^0$C after UV irradiation (Figs. 3, 4).

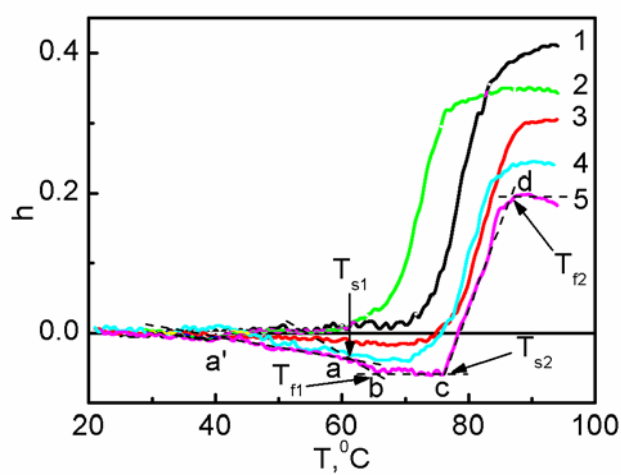

**Fig. 3** The melting curves of DNA without (curves 1 and 2) and with TiO$_2$ NPs (curves 3–5) after 3 hours of the UV irradiation. «1» – [c$_{TiO2}$]=0; «2» – [c$_{TiO2}$]=0 (after 3 hours of the UV irradiation); «3» – [c$_{TiO2}$]=2.5·10$^{-5}$ M; «4» – [c$_{TiO2}$]=10$^{-4}$ M; «5» – [c$_{TiO2}$]=1.5·10$^{-4}$ M; The T$_{s1,2}$ and T$_{f1,2}$ points were obtained using the linear extrapolation (the black dashed lines) of the melting curve segments.

## 4.4 Comparative analysis of $T_m$ and h dependences for UV-irradiated and non-irradiated DNA:TiO$_2$ assemblies

In Figs. 4a-4b we compare the $T_m$ and h-dependences extracted from the irradiated and non-irradiated DNA melting curves in absence and presence of TiO$_2$ NPs as the function of [$c_{TiO2}$]. The $T_m$ dependence shows nonlinear behavior on [$c_{TiO2}$] that may be caused by the complicated processes of the complexation of the biopolymer with NPs. The performed analysis of the DNA melting curve without TiO$_2$ allowed us to determine the thermodynamic parameters of the biopolymer: $T_m$=78.5$^0$C and $h_{m0}$=0.41 (Fig. 3) (where $h_{m0}$ is the value of the hyperchromic coefficient in the absence of TiO$_2$ NPs). The melting temperature of DNA decreases to more than 6$^0$C and the $h_{m0}$ is equal to 0.35 under UV irradiation for three hours (Figs. 4a-4b). This effect may be caused by the partial denaturation of DNA and the appearance of single-stranded fragments under the UV irradiation [11,26]. In the absence of UV irradiation, the addition of [$c_{TiO2}$]=2.5×10$^{-5}$ M to the DNA solution practically does not change the $T_m$ value (Fig. 4a). On the contrary, under UV irradiation, the addition of even [$c_{TiO2}$]=5×10$^{-6}$ M to the DNA solution leads to a sharp increase in the melting temperature by more than 10$^0$C. A further increase in [$c_{TiO2}$] leads to weak changes in the value of this parameter (the changes are within 3$^0$C) for both irradiated and non-irradiated DNA:TiO$_2$ assemblies. The reason for this may be that the $T_m$([$c_{TiO2}$]) dependence is determined by the compensation effects inducing either increasing or decreasing the DNA thermal stability. They are caused by the realization of all possible types of binding of NPs to DNA [16]. The h([$c_{TiO2}$]) dependences extracted from the melting curves for the non-irradiated and irradiated DNA:TiO$_2$ samples are presented in Fig. 4b. In general, both curves demonstrate the similar evolution in the studied concentration range. In the absence of TiO$_2$ NPs, the UV irradiation of DNA solution for 3 hours leads to a 15 % decrease in the h value as compared to non-irradiated DNA. The addition of TiO$_2$ NPs to the DNA solution induces decreasing in the h value, in the absence of UV irradiation. This character of the dependence persists even after a three hour UV irradiation of the DNA:TiO$_2$ complexes.

In the absence of any other conformational transitions, the maximum value of h(T) achieved with complete separation of double-stranded polymer $(h_m)_i$ can be considered as a measure of its degree of helicity ($\Theta$) at T=T$_0$. Analysis of the $(h_m)_0$ evolution shows that in the absence of UV irradiation $(h_m)_0$=0.41, and after three hours of irradiation $(h_m)_0$=0.35, that corresponds to the degree of helicity $\Theta_{T0}$=1. Hence, the degree of DNA helicity in the presence of TiO$_2$ NPs can be determined by the formula: $\Theta=(h_{To}+(h_m)_i)/h_{m0}$. Our estimates show that the degree of helicity of non-irradiated DNA with TiO$_2$ at [$c_{TiO2}$]=1.5·10$^{-4}$M is $\Theta$=(0.12+0.24)/0.41=0.88, and after three hours of UV irradiation $\Theta$=(0.05+0.2)/0.35=0.71. Thus, we can conclude that the addition of TiO$_2$ NPs to the DNA solution leads to a decrease in its degree of the helicity, and UV irradiation is an additional factor that enhances the appearance of the single-stranded unwound regions in DNA.

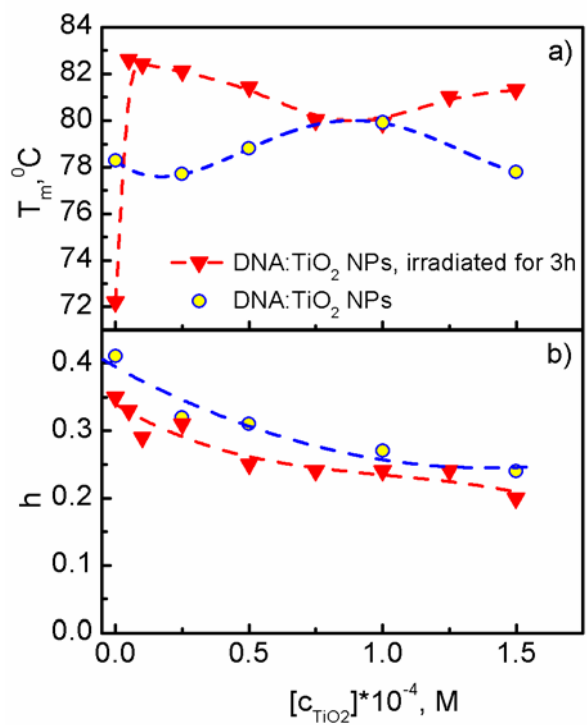

**Fig. 4 (a)** The melting temperature dependence of DNA ($T_m$) as a function of $[c_{TiO2}]$. **(b)** The concentration dependence of hyperchromic coefficient (h) of DNA with $TiO_2$ NPs. The blue and red dashed lines are B-spline interpolations. The size of error bar for $T_m$ and h is smaller than that of the symbols.

## 5 Discussion

According to the available experimental data [16], $TiO_2$ NPs can directly interact with DNA under following scenarios: 1) the electrostatic interaction occurring between positively charged $TiO_2$ NPs and the negatively charged phosphate group of DNA; 2) the interaction of $TiO_2$ NPs with nitrogenous bases in the DNA groove; 3) the formation of the P-O-Ti covalent bonds between the phosphate backbone and the NPs surface. Here we find some similarity in the well known mechanisms of the interaction of the metal ions with DNA [27]. The binding of metal ions to various DNA centers is specific and strongly depends on the nature of the ion. It's known the alkali and alkaline earth metal ions interact mainly with phosphate groups of DNA, and transition metal ions also actively bind with nitrogenous bases. A significant difference is observed in the binding of ions to the native (double-stranded) and denatured (single-stranded) DNA. The density of negative charges on the surface of the double-stranded DNA is much higher than on the single-stranded one. In addition, some active centers on the nitrogenous bases of the native DNA are involved in the formation of the hydrogen bonds and, thus, they are inaccessible for interaction with metal ions. For some ions, the interaction with the N7 atoms of the guanine and adenine is possible. These atoms in the double-stranded structure have the highest molecular electrostatic potential and steric accessibility [19]. Thus, for native DNA, the affinity of ions for phosphate groups is higher than for nitrogenous bases. It was previously found that the binding of metal ions to phosphate groups increases the stability of DNA, while the binding to nitrogenous bases, on the contrary, lowers $T_m$. As a rule, the character of the dependence of $T_m$ on the ion concentration is determined by the difference in the binding of ions with phosphates and nitrogenous bases of DNA [27]. According to our experimental data, along with the electrostatic interaction of the positively charged $TiO_2$ NPs with the negatively charged phosphate DNA groups, there is also a binding of $TiO_2$ NPs with the N7 atoms of guanine of nitrogenous bases.

The argument in favor of this assumption is the data presented in Table A. The analysis of the data shows that the narrowing of the DNA melting interval ($\Delta T=T_f-T_s$) is observed in the absence of irradiation in the concentration range of $[c_{TiO2}]=2.5*10^{-5}$-$5*10^{-5}$ M and $[c_{TiO2}]=10^{-4}$-$1.5 \cdot 10^{-4}$ M. After a three-hour irradiation, the narrowing of the melting interval is observed in the range of $[c_{TiO2}]=5 \cdot 10^{-6}$-$7.5 \cdot 10^{-5}$ M and $[c_{TiO2}]=1.25*10^{-4}$-$1.5*10^{-4}$ M (Table B). According to the data presented in Tables A and B, the narrowing of $\Delta T$ is mainly caused by the predominant decrease in the temperature of the melting end, which is determined by the stability of the DNA blocks enriched in GC pairs. The expansion of the melting interval is observed in the concentration range of $[c_{TiO2}]=5 \cdot 10^{-5}$-$10^{-4}$M for the non-irradiated DNA and $[c_{TiO2}]=5 \cdot 10^{-5}$-$1.25 \cdot 10^{-4}$M for the irradiated DNA. This effect is apparently due to the predominant increase in the temperature of the DNA melting end. The possibility of the interaction of NPs with the N7 guanine atoms in the DNA groove was shown for the example of DNA binding to AgNPs and AuNPs [18].

**Table A.** The concentration dependences of the melting range ($\Delta T$) and changes in the temperatures of the beginning ($T_s$) and end ($T_f$) of the helix-coil transition of the non-irradiated DNA.

| $[c_{TiO2}]$, M | $T_s$,°C | $T_f$,°C | $\Delta T$,°C |
|---|---|---|---|
| $2.5 \cdot 10^{-5}$ | 73.1 | 82.3 | 9.2 |
| $5 \cdot 10^{-5}$ | 75.5 | 82.2 | 6.7 |
| $10^{-4}$ | 74.1 | 85.5 | 11.4 |
| $1.5 \cdot 10^{-4}$ | 74.1 | 81.6 | 7.5 |

**Table B.** The concentration dependences of the melting range (ΔT) and changes in the temperatures of the beginning ($T_s$) and end ($T_f$) of the helix-coil transition of the irradiated DNA.

| [$c_{TiO2}$], M | $T_s$, °C | $T_f$, °C | ΔT, °C |
|---|---|---|---|
| 5·10⁻⁶ | 76.8 | 88.4 | 11.6 |
| 10⁻⁵ | 77.5 | 87.4 | 9.9 |
| 2.5·10⁻⁵ | 77.2 | 86.9 | 9.7 |
| 5·10⁻⁵ | 77.3 | 85.4 | 8,1 |
| 7.5·10⁻⁵ | 77 | 83 | 6 |
| 10⁻⁴ | 75.6 | 84 | 8.4 |
| 1.25·10⁻⁴ | 76.2 | 86.1 | 9.9 |
| 1.5·10⁻⁴ | 76.1 | 85.7 | 9.6 |

In the view of the above, the difference in the $T_m$([$c_{TiO2}$]) dependences (Fig. 4a) for irradiated and non-irradiated DNA:TiO$_2$ NPs assemblies is apparently due to the different types of the interactions between DNA and TiO$_2$ NPs. The decrease in $T_m$ with the increase in [$c_{TiO2}$] may be due to the fact that the main contribution to the change in $T_m$ introduces the interaction of NPs with the N7 guanine atoms in the DNA groove. The increase in the thermal stability of the biopolymer is primarily due to the electrostatic interaction of positively charged NPs with negatively charged DNA phosphate groups. In addition, as mentioned above, the interaction of TiO$_2$ NPs with DNA leads to a decrease in the h value (Fig. 4b). In most cases, this effect is a consequence of the base stacking disruption and it's enhanced with an increase in the TiO$_2$ NPs concentration.

## 6 Conclusions

The effect of TiO$_2$ NPs on the structural stability of DNA was studied using optical absorption spectroscopy with DNA thermal denaturation in the range of 20-94°C. It was shown that upon DNA heating from about 25 to 44°C the stabilization of DNA duplex occurs in the presence of TiO$_2$ NPs. This additional DNA stabilization indicates that partial DNA unwinding appears as a result of the direct binding of the biopolymer to NPs. DNA binding with TiO$_2$ NPs is manifested in the change of the DNA $T_m$ and decreasing the hyperchromic coefficient that decreases with increasing NPs concentration. These changes of the parameters of DNA stability indicate the partial biopolymer duplex unwinding.

It was found that UV irradiation of DNA for 3 hours leads to the decrease in the melting temperature by more than 6°C and the decrease in the h value. This effect can be caused by partial DNA duplex unwinding. The NPs injection to the biopolymer solution induced the additional effect on the DNA thermal stability at room temperature ($T_m$ increases by about $10^0 C$ ) and decrease in the h value under UV irradiation.

The data obtained in this work allow us to make the following conclusion that the adding of $TiO_2$ NPs in the buffer solution with DNA leads to partial DNA duplex unwinding. UV irradiation enhances the toxic effect of these NPs. The consequence of this $TiO_2$ NPs effect on the living organism can manifest itself in a disruption of the normal functioning of the genetic apparatus of the cell, as well as mutagenesis and carcinogenesis. The obtained results can be also used in the creation of self-cleaning antibacterial surfaces, as well as in medicine.

**Acknowledgments**

Authors acknowledge financial support from National Academy of Sciences of Ukraine (Grant No. 0120U100157).

## References

1. S. Kwon, M. Fan, A.T. Cooper, Crit. Rev. Env. Sci. Tec. **38**(3), 197 (2008)
2. Ch. O. Robichaud, A. E. Uyar, M. R. Darby, L. G. Zucker, M. R. Wiesner, Environ. Sci. Technol. **43**(12), 4227 (2009)
3. J. Besley, V. Kramer, S. Priest, J. Nanoparticle Res. **10**(4), 549 (2008).
4. Engineered nanomaterials: a review of the toxicology and health hazards. Safe work, Australia, November, 143 p. (2009)
5. S. Patel, P. Patel, S. R. Bakshi, Cytotechnology **69**(2), 245 (2017)
6. Characterising the potential risks posed by engineered nanoparticles: A second UK government research report. Department of Environment, Food and Rural Affairs (DEFRA, London, UK, 2007).
7. EU nanotechnology R&D in the field of health and environmental impact of nanoparticles (Compiled by Pilar Aguar and José Juan Murcia Nicolás, Unit G4 Nano and Converging Sciences and Technologies European Commission, Research DG, 2008).
8. V. Aruoja, H.C. Dubourguier, K. Kasemets, A. Kahru, Sci. Total Environ. **407**(4), 1461 (2009)
9. I. Velzeboer, A. J. Hendriks, A. M. J. Ragas, D. Van de meent, Environ. Toxicol. and Chem. **27**(9), 1942 (2008)
10. C. Blaise, F. Gagne, J. F. Ferard, P. Eullaffroy, Environ. Toxicol. **23**, 591 (2008)

11. R. P. Rastogi, Richa, A. Kumar, M. B. Tyagi, R. P. Sinha, J. Nucleic Acids **2010**, 592980 (2010)
12. N. González, Maria del Àngels Custal, D. Rodríguez, Jordi-Roger Riba, E. Armelin, Materials Research. **20**(4), 1082 (2017)
13. T. A. Egerton, Molecules **19**, 18192 (2014)
14. X. Zhang, F. Wang, B. Liu, E. Y. Kelly, M. R. Servos, J. Liu, Langmuir **30**(3), 839 (2014)
15. K. Li, S. Du, S. Van Ginkel, Y. Chen, *Atomic Force Microscopy Study of the Interaction of DNA and Nanoparticles*, Nanomaterial, Advances in Experimental Medicine and Biology, edited by D.G. Capco, Y. Chen, Vol. **811** (Springer Science+Business Media Dordrecht, 2014), pp. 93-109.
16. S. Patel, P. Patel, S. B. Undre, Sh. R. Pandya, M. Singh, S. Bakshi, Journal of Molecular Liquids **213**, 304 (2016)
17. K. S. El-said, E. M. Ali, K. Kanehira, A. Taniguchi, J. Nanobiotechnol. **12**, 48 (2014)
18. N. A. Kasyanenko, A. A. Andreeva, A. V. Baryshev, V. M. Bakulev, M. N. Likhodeeva, P. N. Vorontsov-Velyaminov, J. Phys. Chem. B **123**, 9557 (2019)
19. V. A. Sorokin, V. A. Valeev, E. L. Usenko, V. V. Andrushchenko, Int. J. Biol. Macromol. **50**(3), 854 (2012)
20. K. Higasi, H. Baba, A. Rembaum, *Quantum Organic Chemistry* (Interscience Publishers A division of John Wiley & sons, New York, 1965)
21. E.L. Usenko, V.A. Valeev, A.Yu. Glamazda, V.A. Karachevtsev, J. Spectrosc. Article ID 8850214 (7) (2020).
22. A. P. Sarapultsev, S. V. Rempel, Ju. V. Kuznetsova, G. P. Sarapultsev, J. Ural Med. Acad. Sci. (3), 97 (2016)
23. N. S. Leonenko, O. B. Leonenko, Innov. Biosyst. Bioeng. **4**(2), 75 (2020)
24. H. M. Ali, H. Babar, T. R. Shah, M. U. Sajid, M. A. Qasim, S. Javed, Appl. Sci. **8**(4), 587 (2018)
25. Th. Dasri, P. Chaiyachate, S. Audtarat, P. Charee, A. Chingsungnoen, J. Met. Mater. Miner. **30**(3), 30 (2020)
26. J. Kiefer, *Effects of Ultraviolet Radiation on DNA*, Chapter from book Chromosomal alterations: Methods, results and importance in human health, edited by Günter Obe and Vijayalaxmi, (Springer-Verlag Berlin Heidelberg, 2007), pp.39-53.
27. Yu. P. Blagoi, V. L. Galkin, G. O. Gladchenko, S. V. Kornilova, V. A. Sorokin, A. G. Shkorbatov, *Metal Complexes of Nucleic Acids in Solutions* (Naukova Dumka, Kiev, 1991) [in Russian].